\documentclass[final,5p,times,twocolumn]{elsarticle}

\usepackage{amssymb}
\usepackage{lipsum}
\usepackage{amsmath}
\usepackage{xcolor}
\usepackage{multirow}
\usepackage{soul}
\usepackage{physics}
\usepackage[numbers]{natbib}
\usepackage{caption}
\usepackage{subcaption}
\usepackage{graphicx}
\usepackage{fontspec}
\usepackage{xspace}
\usepackage{tabularx}
\usepackage{makecell}
\usepackage{booktabs}
\usepackage{url}
\usepackage{hyperref}
\usepackage{subfiles}
\usepackage{etoolbox}

\pretocmd{\MaketitleBox}{\vspace*{-6mm}}{}{}

\hypersetup{hidelinks}
\newcommand{\software}[1]{\texttt{#1}}

\newcommand{\mg}{\software{MadGraph5\_aMC@NLO (v3.2.0)}\xspace}
\newcommand{\pythia}{\software{PYTHIA8 (v8.311)}\xspace}
\newcommand{\pythiaShort}{\software{PYTHIA8}\xspace}
\newcommand{\delphes}{\software{DELPHES (v3.5.0)}\xspace}
\newcommand{\delphesShort}{\software{DELPHES}\xspace}
\newcommand{\combine}{\software{Combine (v10.4.2)}\xspace}
\newcommand{\combineShort}{\software{Combine}\xspace}

\journal{Physics Letters B}

\begin{document}

\begin{frontmatter}

%% Title, authors and addresses

%% use the tnoteref command within \title for footnotes;
%% use the tnotetext command for theassociated footnote;
%% use the fnref command within \author or \affiliation for footnotes;
%% use the fntext command for theassociated footnote;
%% use the corref command within \author for corresponding author footnotes;
%% use the cortext command for theassociated footnote;
%% use the ead command for the email address,
%% and the form \ead[url] for the home page:
%% \title{Title\tnoteref{label1}}
%% \tnotetext[label1]{}
%% \author{Name\corref{cor1}\fnref{label2}}
%% \ead{email address}
%% \ead[url]{home page}
%% \fntext[label2]{}
%% \cortext[cor1]{}
%% \affiliation{organization={},
%%            addressline={}, 
%%            city={},
%%            postcode={}, 
%%            state={},
%%            country={}}
%% \fntext[label3]{}

\title{Reinterpreting ATLAS Dijet and Supersymmetry Searches in A Baryon-Number-Violating Heavy Scalar Framework}

%% use optional labels to link authors explicitly to addresses:
%% \author[label1,label2]{}
%% \affiliation[label1]{organization={},
%%             addressline={},
%%             city={},
%%             postcode={},
%%             state={},
%%             country={}}
%%
%% \affiliation[label2]{organization={},
%%             addressline={},
%%             city={},
%%             postcode={},
%%             state={},
%%             country={}}
%\author[Tamu]{Antimo Cagnotta}
\author[VB]{Tianyi Yang}
\author[VB]{Denis Rathjens}
\author[VB]{Alfredo Gurrola}
\author[VB]{Francesco Romeo}
\affiliation[VB]{organization={Vanderbilt University},%Department and Organization
            addressline={2301 Vanderbilt Place}, 
            city={Nashville},
            postcode={37235}, 
            state={Tennessee},
            country={U.S.A.}}
            
\begin{abstract}
A reinterpretation of ATLAS searches targeting $b$-tagged dijet resonances and monotop-alike signatures at a center-of-mass energy of 13 TeV is presented using $pp$ collision data corresponding to integrated luminosities of 139 $\mathrm{fb}^{-1}$.
These searches are used to constrain a baryon-number-violating new physics model that introduces two scalar mediators $X_{1}$ and $X_{2}$ to simultaneously address the baryon asymmetry of the universe and the dark matter relic abundance.
By exploiting heavy-flavor signatures and allowing for jet-rich final states, this reinterpretation provides complementary sensitivity to the model relative to non-flavor-tagged inclusive dijet searches and conventional monotop reinterpretations.
Assuming the mediators decay either into a pair of down-type quarks or into a top quark and a dark matter candidate, we translate the experimental upper limits into the model's parameter space.
95\% confidence level exclusion limits are derived in the $\lambda_1$-$\lambda_2$ coupling plane for benchmark mediator masses of 1.5, 1.75, and 2.0 TeV.
For all three benchmark masses, the $b$-tagged dijet reinterpretation excludes $\lambda_1 > 0.45$, while the bottom-squark reinterpretation excludes the region with $\lambda_1 > 0.3$ and $\lambda_2 > 0.6$, within the respective validated parameter domains.
\end{abstract}

%%Graphical abstract
%\begin{graphicalabstract}
%\includegraphics{grabs}
%\end{graphicalabstract}

%%Research highlights
%\begin{highlights}
%\item Research highlight 1
%\item Research highlight 2
%\end{highlights}

\begin{keyword}
%% keywords here, in the form: keyword \sep keyword, up to a maximum of 6 keywords
Beyond the Standard Model (BSM) \sep Scalar Mediators \sep Collider Phenomenology \sep LHC Reinterpretation

%% PACS codes here, in the form: \PACS code \sep code

%% MSC codes here, in the form: \MSC code \sep code
%% or \MSC[2008] code \sep code (2000 is the default)

\end{keyword}

\end{frontmatter}

%\tableofcontents

%% \linenumbers

%% main text
\section{Introduction}

The observed matter–antimatter asymmetry in the universe \cite{Canetti:2012zc}, alongside the established relic abundance of dark matter (DM) \cite{Planck:2018vyg}, provides strong motivation for extensions of the Standard Model (SM) that incorporate baryon number violation (BNV).
Such models predict novel collider signatures that can be probed through reinterpretation of existing experimental searches.
Historically, heavy scalar mediators in BNV frameworks have been constrained primarily through inclusive dijet resonance searches or standard monotop analyses.
However, models featuring preferential couplings to third-generation quarks offer unique phenomenological opportunities that remain underexplored.
Therefore, we demonstrate in this study that leveraging analyses beyond these traditional approaches (e.g. ATLAS searches for $b$-tagged dijet resonances and supersymmetry \cite{ATLAS:2019fgd, ATLAS:2021jyv, ATLAS:2021yij}) provides a meaningful and previously unexploited reach into the BNV parameter space.
Analogous CMS searches, including $b$-tagged dijet resonance \cite{CMS:2022eud} and inclusive dijet resonance \cite{CMS:2019gwf} searches, could likewise be reinterpreted to constrain the BNV parameter space.
However, we restrict our reinterpretation to ATLAS analyses in this work.
In addition to these reinterpretations, we perform an exploratory study of a heavy-flavor-enriched dijet category requiring an additional subleading $b$-tagged jet, to assess whether such a dedicated selection could further improve sensitivity to the model.

In this work, we focus on a $\Delta B = 2$ BNV model \cite{Kalsi:2024hfe} featuring new scalar mediators $X_{1}$ and $X_{2}$.
This model predicts that these mediators can decay either to a top quark and a stable, neutral fermion $\psi$; or to down-type (anti)quark pairs ($db$, $sb$, and their charge-conjugate states).
By virtue of its stability and weak interactions, $\psi$ serves as a viable DM candidate.
The phenomenology of this framework is largely governed by two coupling parameters: $\lambda_1$, which controls the interaction between the mediator, a bottom quark, and a lighter down-type quark ($d$ or $s$); and $\lambda_2$, which governs the coupling of the mediator to the top quark and $\psi$.
A detailed description of the model, including the interaction Lagrangian and its parameters, is presented in Section~\ref{sec:modeldef}.

To evaluate the viability of this model, we derive constraints on the BNV model parameter space by generating Monte Carlo (MC) simulated events with relevant $b$-tagged dijet and single-top-quark plus $p_T^{\mathrm{miss}}$ topologies.
The signal event generation and processing are outlined in Section~\ref{sec:simulation}.

The monotop-alike signatures we consider include both hadronic and leptonic top decays, but the signal acceptance in the leptonic channel for \cite{ATLAS:2021jyv} turns out to be insufficient to exclude any parameter space.
Consequently, the exclusion bounds we present are derived from \cite{ATLAS:2019fgd} impacting the down-type quark decay mode of $X$ and  \cite{ATLAS:2021yij} constraining the $t\psi$ decay mode.
The analysis methodology is detailed in Section~\ref{sec:methodology}, followed by the presentation of the resulting exclusion limits in the $\lambda_1$-$\lambda_2$ plane in Section~\ref{sec:results}.
Finally, our conclusions are summarized in Section~\ref{sec:summary}.

\section{Model Definition and Parameter Space} \label{sec:modeldef}

The $\Delta B = 2$ BNV model considered in this study introduces two charged scalar fields $X_{\alpha}$ ($\alpha = 1,2$), which transform as color triplets under $SU(3)_{c}$, as singlets under $SU(2)_{L}$, and carry a fractional hypercharge $+4/3$.
These scalars mediate effective interactions between quarks and a fermionic DM candidate, $\psi$.
Such operators can facilitate baryogenesis through the interference of tree-level and loop-level processes in decays of the type $X_{\alpha} \to \bar{d}_{i} \bar{d}_{j}$.
The minimal interaction Lagrangian in the Weyl basis is given by
\begin{equation}
    \mathcal{L} \supset \lambda_{\alpha i} X_{\alpha} \psi u_{i}^{c} + \lambda'_{\alpha ij} X_{\alpha}^{*} d_{i}^{c} d_{j}^{c} + \frac{m_{\psi}}{2} \bar{\psi}^{c} \psi + \text{h.c.}
\end{equation}
where $i, j$ denote quark flavor indices ($i \neq j$ due to $SU(3)_c$ antisymmetry) and $\lambda_{\alpha i}, \lambda'_{\alpha ij}$ are the relevant coupling constants.
Hereafter, this framework is referred to as the mono-$X$ model.

To simplify the phenomenology for collider reinterpretation, we consider the limit $m_{X_1} \ll m_{X_2}$.
In this limit, the production of $X_{2}$ is kinematically suppressed, making its collider signatures effectively subdominant relative to those of $X_{1}$.
While optimal baryogenesis scenarios typically favor $m_{X_1} \simeq m_{X_2}$, previous studies \cite{Kalsi:2024hfe} claimed that a viable baryon asymmetry can still be generated in this limit.
Therefore, we focus exclusively on the lighter mediator $X_1 \equiv X$ and its parameter space, which retains a meaningful, though non-ideal, connection to the baryogenesis-motivated setup.

\begin{figure}[!hbpt]
  \centering
  \includegraphics[width=0.6\columnwidth]{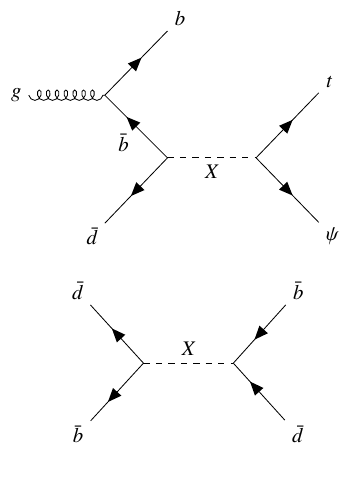}
  % % --- Top subfigure ---
  % \begin{subfigure}{0.6\columnwidth}  % or \textwidth if you want full width
  %   \centering
  %   \includegraphics[width=\linewidth]{images/feynman-top.pdf}
  %   \label{fig:X_feynman_a}
  % \end{subfigure}

  % \vspace{1.5em} % optional vertical space between the two

  % % --- Bottom subfigure ---
  % \begin{subfigure}{0.6\columnwidth}
  %   \centering
  %   \includegraphics[width=\linewidth]{images/feynman-bottom.pdf}
  %   \label{fig:X_feynman_b}
  % \end{subfigure}

  \caption{Representative Feynman diagrams illustrating production and decay topologies in the Mono-$X$ model. The upper diagram shows the decay of the scalar mediator $X$ into a top quark and a fermionic DM candidate $\psi$, while the lower diagram depicts the hadronic decay of $X$ into a bottom antiquark and a lighter down-type antiquark. Additional jet activity from bottom-associated production via initial-state gluon splitting is included in the upper diagram to yield final-state topologies compatible with those targeted by ATLAS searches.}
  \label{fig:X_feynman}
\end{figure}

The phenomenology of $X$ is dictated by its mass, $m_X$, its decay width, and its Yukawa-like couplings to the SM fermions and DM.
For the signal simulation in this study, the DM candidate mass is fixed to $1~\text{GeV}$, and the total decay width of the mediator is evaluated dynamically as a function of the masses and couplings using the ``auto'' width setting in \mg \cite{Alwall:2014hca}.

To map the model onto experimental signatures, we define two simplified coupling parameters:
\begin{itemize}
    \item $\lambda_1$: The universal coupling of $X$ to a bottom quark and a lighter down-type quark ($d$ or $s$), mapping to $\lambda'_{131} = \lambda'_{132} \equiv \lambda_1$. This coupling primarily governs the decay of the mediator into a pair of down-type quarks, resulting in $b$-tagged dijet signatures.
    \item $\lambda_2$: The coupling of $X$ to the top quark and the DM particle, mapping to $\lambda_{13} \equiv \lambda_2$. This coupling governs the decay into a top quark and a stable $\psi$ fermion, manifesting as a monotop-alike signature characterized by a top quark and missing transverse momentum ($p_{\text{T}}^{\text{miss}}$).
\end{itemize}

The scalar $X$ is preferentially produced in association with a $b$-quark originating from an initial-state gluon splitting (e.g., $g \bar{d} \to X b$).
This production mechanism naturally introduces the additional hard jets in the final state required to match the event topologies and selection criteria targeted by the reinterpreted ATLAS analyses.
Representative Feynman diagrams illustrating these production and decay topologies are shown in Figure \ref{fig:X_feynman}.

Because the production mechanism relies on the coupling to down-type quarks, the total production cross section scales quadratically with the primary coupling, $\sigma \propto \lambda_1^2$.
Neglecting phase space differences from final-state particle masses, the branching fractions ($\text{BR}$) for the competing decay channels can be expressed analytically as functions of the two coupling parameters:
\begin{gather*}
\mathrm{BF} \left( X \to \bar{d}\bar{b} \right) + \mathrm{BF} \left( X \to \bar{s}\bar{b} \right) = \frac{4\lambda_{1}^{2}}{4\lambda_{1}^{2} + \lambda_{2}^{2}} \\
\mathrm{BF} \left( X \to t \psi \right) = \frac{\lambda_{2}^{2}}{4\lambda_{1}^{2} + \lambda_{2}^{2}}
\end{gather*}

These relations form the basis of the semi-analytical scaling used to map the experimental cross-section limits continuously onto the $(\lambda_1, \lambda_2)$ parameter plane.

\section{Simulation} \label{sec:simulation}

MC event generation is performed at leading order using \mg \cite{Alwall:2014hca}.
We simulate the bottom-associated production of the scalar mediator at the matrix-element level, generating inclusive samples of $p p \to X$ and $p p \to X + 1j$ at leading order in QCD.
We consider either $X \to t\psi$ or $X \to \bar{d}\bar{b},\,\bar{s}\bar{b}$ decays, depending on the final state targeted by the analysis being reinterpreted.
A five-flavor scheme is employed using the NNPDF2.3LO parton distribution function (PDF) set \cite{2013290} to allow for initial-state $b$-quarks.
To avoid double-counting of jet radiation between the matrix-element calculation and the subsequent parton shower, the MLM matching scheme \cite{Alwall:2007fs, Mangano:2006rw} is applied.
Following the parton-level generation, events are interfaced with \pythia \cite{Sjostrand:2014zea} using the Monash 2013 tune \cite{Skands:2014pea} to simulate parton showering, hadronization, and the underlying event.

For the exploratory study of the additional-$b$-tag dijet selection, QCD multijet background samples are also generated with \mg in the five-flavor scheme.
Matrix-element configurations with two, three, and four final-state jets are included in order to model additional jet activity relevant to the proposed selection.
To obtain sufficient statistics over the steeply falling multijet spectrum, the samples are generated in exclusive generator-level $H_T$ intervals of $[500,700]$, $[700,900]$, $[900,1200]$, $[1200,1600]$, $[1600,2200]$, $[2200,3000]$, and $[3000,\infty)$ GeV.
The generator-level $H_T$ requirement is imposed through the \mg \texttt{ihtmin}/\texttt{ihtmax} parameters, with $H_T$ defined as the scalar sum of the transverse momenta of the generated partons.
The resulting samples are subsequently showered and passed through the same detector simulation and event selection used for the $b$-tagged dijet study, and the different $H_T$ intervals are combined according to their respective generated cross sections.

Detector effects are modeled using the \delphes fast-simulation framework \cite{deFavereau:2013fsa}.
Because the experimental acceptances vary significantly between the targeted analyses, custom \delphesShort cards are tailored for each reinterpretation where applicable.
Specifically, the jet clustering radius parameter ($R$) and kinematic $p_{\text{T}}$ thresholds are modified to match each respective ATLAS search.
For the $b$-tagged dijet reinterpretation~\cite{ATLAS:2019fgd}, ATLAS explicitly notes that the $b$-tagging performance depends heavily on the jet $p_{\text{T}}$.
To account for this dependence, we construct $p_{\text{T}}$-dependent $b$-tagging efficiencies and misidentification rates by interpolating between the performance values reported by ATLAS.
The simulation-to-data scale factors provided in the same analysis are also applied multiplicatively to the corresponding $b$-tagging efficiencies in each $p_{\text{T}}$ interval.
Since the dijet search does not report the tagging performance for $p_{\text{T}} \leq 500~\text{GeV}$, which is relevant when identifying additional subleading $b$-tagged jets in the exploratory study, we supplement this parameterization with the performance results in~\cite{ATLAS:2022qxm}, which reports the same tagging algorithm over $20 \leq p_{\text{T}} \leq 250~\text{GeV}$.
The resulting low- and high-$p_{\text{T}}$ measurements are combined into a continuous approximate efficiency parameterization in the corresponding \delphesShort card.
For the remaining object-identification efficiencies, including $b$-tagging and $\tau$-tagging where applicable, the values reported by the corresponding ATLAS analyses are invoked as constants in the \delphesShort cards.

\section{Methodology Overview} \label{sec:methodology}

In order to reinterpret ATLAS searches in the context of the mono-$X$ model, we connect the experimental and theoretical inputs via the cross section, where a similar enough model cross section exclusion is available, or estimate the signal yield by implementing the analysis selections.

The choice of ATLAS analyses used in this reinterpretation is guided by the characteristic heavy-flavor and jet-rich signatures of the mono-$X$ model.
In the dijet channel, the mediator decay involves a bottom quark together with a light down-type quark, motivating the use of a $b$-tagged dijet resonance search rather than an inclusive dijet search.
The $b$-tag requirement exploits a distinctive feature of the signal topology and helps reduce the otherwise large SM QCD multijet background.
For the $X \to t\psi$ decay mode, the resulting final states can contain additional jet activities from top-quark decay and associated radiation.
We therefore reinterpret a supersymmetry-motivated bottom-squark search with a broader jet multiplicity requirement, $N_{\mathrm{jets}} \in [2,4]$, which is well matched to these jet-rich configurations.
These selections provide sensitivity to complementary regions of the mono-$X$ parameter space relative to non-flavor-tagged dijet searches and more restrictive conventional monotop selections.
A summary of the reinterpreted ATLAS searches, including their targeted signatures, integrated luminosities, and specific recasting strategies, is provided in Table~\ref{tab:atlas_summary}.

\begin{table*}[!hpbt]
\begin{center}
\begin{small}
\begin{tabular}{cccccc}
\toprule
\textbf{ATLAS Search} & $N_{\mathrm{jets}}$ & \textbf{Other Criteria} & \textbf{Luminosity ($\text{fb}^{-1}$)} & \textbf{Reinterpretation Method} & \textbf{SR Name} \\
\midrule
Dijet Resonance \cite{ATLAS:2019fgd} & $[2, +\infty)$ & $\ge 1 \, b$ & 139 & \delphesShort acceptance with $b^*$ limits & $\ge 1$ $b$-tag\\
Bottom squark \cite{ATLAS:2021yij} & $[2, 4]$ & $2 \, b$, $p_{\text{T}}^{\text{miss}}$ & 139 & \delphesShort simulation with \combineShort & SRA\\
Top squark \cite{ATLAS:2021jyv} & $[2, +\infty)$ & $\geq 2 \, b$, $1 \, \tau$, $p_{\text{T}}^{\text{miss}}$ & 139 & \makecell{\delphesShort simulation with \\ Single-tau limits} & Single-tau\\
\bottomrule
\end{tabular}
\end{small}
\end{center}
\caption{Summary of the reinterpreted ATLAS searches, detailing the targeted experimental final states, the corresponding Run 2 integrated luminosities, and the specific recasting methodologies implemented in this framework.}
\label{tab:atlas_summary}
\vskip -0.1in
\end{table*}

When an analysis is reinterpreted using \delphesShort simulation, the selection is applied to the reconstructed events, which are then weighed by the product of cross section, branching fraction, and integrated luminosity, to estimate the predicted yield.
This yield is then compared against the 95\% confidence level (CL) upper limit on the number of signal events (or equivalently the model-independent visible cross section) reported by ATLAS.
A given parameter point in the mono-$X$ model is considered excluded if the predicted signal yield exceeds the observed upper limit.

For the bottom-squark reinterpretation~\cite{ATLAS:2021yij}, we employ the \combine statistical framework~\cite{CMS:2024onh} to obtain a combined 95\% CL limit from the five ATLAS signal regions \texttt{SRAmct250i}, \texttt{SRAmct350i}, \texttt{SRAmct450i}, \texttt{SRAmct550i}, and \texttt{SRAmct650i}.
The predicted signal yields, obtained from the \delphesShort simulated samples and normalized to the integrated luminosity, enter a combined counting-experiment likelihood.
Three sources of uncertainty are included through log-normal nuisance parameters.
For each region $i$, the quoted SM background expectation $B_i \pm \Delta B_i$ is assigned a multiplicative uncertainty factor of $1 + \Delta B_i / B_i$.
The statistical uncertainty from the finite number $N_{\mathrm{pass},i}$ of selected simulated signal events is represented by a factor of $1 + 1 / \sqrt{N_{\mathrm{pass},i}}$.
In this approximate likelihood, the background uncertainties and signal statistical uncertainties are treated as independent between regions and from one another.
Finally, a flat 10\% uncertainty is assigned to the \mg signal cross section and treated as fully correlated across all five regions, motivated by the accuracy criterion for the semi-analytical scaling described below.

To translate the results into exclusion bounds on the mono-$X$ model parameter space, we perform a systematic scan over the mediator mass $m_{X}$ and the coupling strengths $\lambda_{1}$ and $\lambda_{2}$.
For benchmark mediator masses of 1.5, 1.75, and 2.0 TeV, the couplings $\lambda_{1}$ and $\lambda_{2}$ are varied independently over the range 0.1 to 1.0 in steps of 0.1, forming a grid.
Additional simulation points outside the nominal scan were generated to examine finite-width effects, including $(\lambda_1, \lambda_2) = (1.3, 0.9)$ at $m_X = 1.5~\mathrm{TeV}$.
At each grid point, the corresponding production cross sections are obtained from \mg simulation interfaced with \pythiaShort.
To achieve a continuous mapping of the parameter space between the simulated grid points, we apply a semi-analytical scaling.
The scaling coefficient, $c$, is determined by fitting the analytical expression $\sigma_{\mathrm{theory}} = c \lambda_1^2 \times \text{BF}(\lambda_1, \lambda_2)$ to the \mg simulated cross sections across the generated grid.

At large values of $\lambda_1$ or $\lambda_2$, the total decay width of the mediator increases substantially, causing the simulated cross sections to deviate from the narrow-width scaling behavior due to off-shell and large-coupling effects.
To ensure the integrity of our reinterpretation, we define a validated scaling region within which the semi-analytical prediction accurately reproduces the \mg simulation, with a relative difference of $\leq 10\%$.
Specifically, for each probed parameter point, we require:
\begin{equation}
    \frac{\left\lvert  \sigma_{\mathrm{theory}} - \sigma_{\mathrm{MG}} \right\rvert}{\sigma_{\mathrm{MG}}} \leq 10\%
\end{equation}
Within this domain, the 10\% uncertainty on the \mg signal cross section described above is included in the statistical analysis.
Outside of this boundary, the semi-analytical framework is excluded from the final constraint curves presented in Section~\ref{sec:results}.
Within the bounded parameter space, predictions are compared to the derived ATLAS exclusion limits, allowing us to compute continuous 95\% CL exclusion contours in the $\lambda_1$-$\lambda_2$ plane for each mediator mass.

\section{Results} \label{sec:results}

In this section, we present the results of reinterpreting ATLAS searches for $b$-tagged dijet resonances and bottom squarks \cite{ATLAS:2019fgd, ATLAS:2021yij}, in the context of the Mono-$X$ model.
The analyses considered target complementary final states that arise from different mediator decay modes: dijet resonances with $b$-tagged jets, multijet final states with $b$-jets and missing transverse momentum ($p_{\text{T}}^{\text{miss}}$), and events containing $b$-jets, $\tau$ leptons, and $p_{\text{T}}^{\text{miss}}$.
Each search probes a distinct experimental signature of the model, providing insight into the constraints on the $(\lambda_{1}, \lambda_{2})$ parameter space for different mediator masses.
The derived exclusion contours for the successful reinterpretations are presented sequentially, followed by a discussion on the methodological limitations encountered in the $\tau$ lepton channel.

\begin{figure*}[!hbpt]
  \centering

  \begin{subfigure}[t]{0.33\textwidth}
    \centering
    \includegraphics[width=\linewidth]{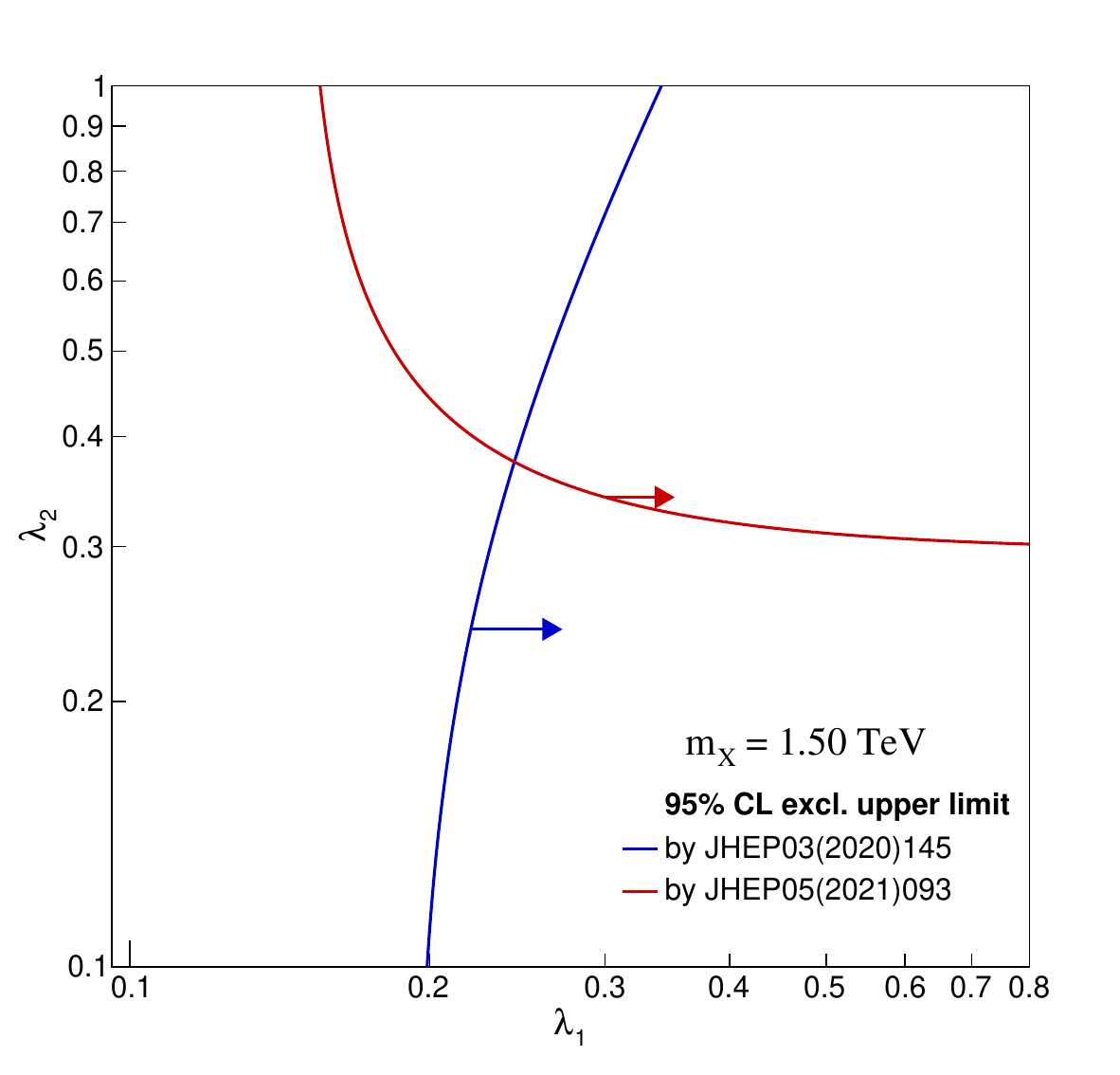}
  \end{subfigure}\hfill
  \begin{subfigure}[t]{0.33\textwidth}
    \centering
    \includegraphics[width=\linewidth]{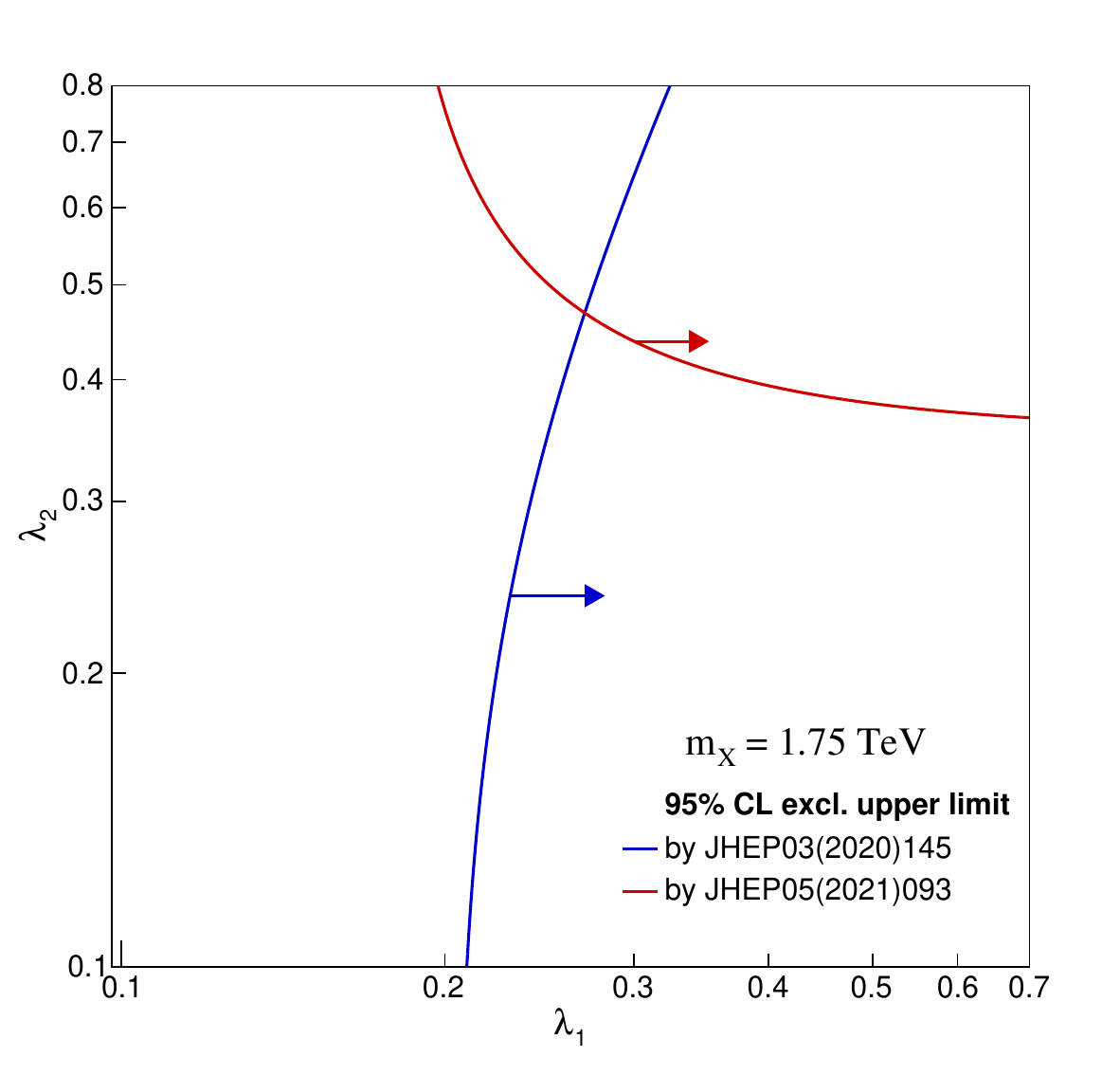}
  \end{subfigure}\hfill
  \begin{subfigure}[t]{0.33\textwidth}
    \centering
    \includegraphics[width=\linewidth]{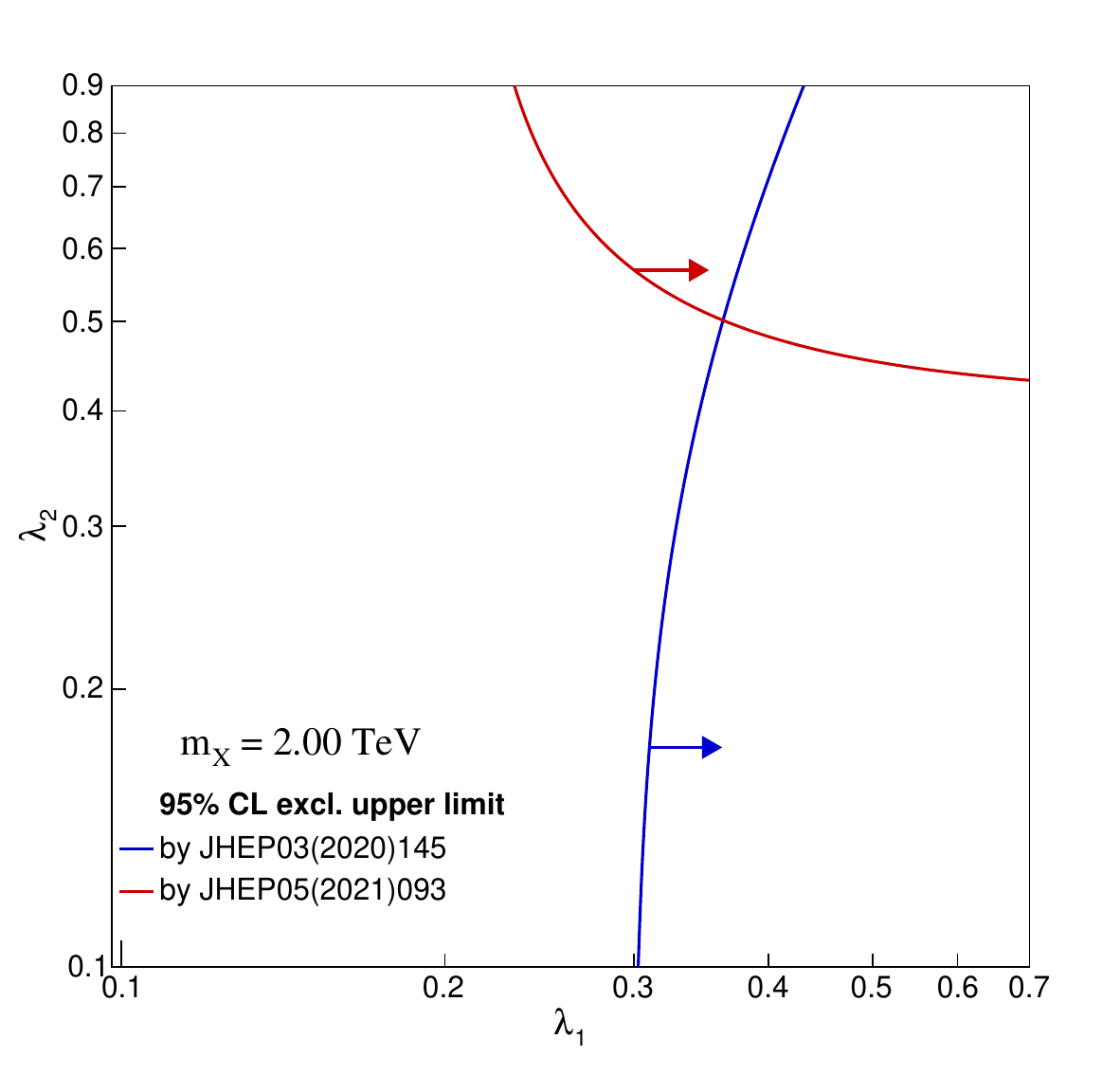}
  \end{subfigure}

  \caption{Exclusion contours in the $(\lambda_{1}, \lambda_{2})$ parameter space for the mono-$X$ model with mediator masses $m_{X} = 1.5$, $1.75$, and $2.0~\text{TeV}$ (left to right, respectively), derived from the reinterpretation of ATLAS searches. The solid blue curves represent the 95\% CL exclusion limits from the $b$-tagged dijet resonance search, while the solid red curves delineate the limits from the bottom squark search. For each channel, the validated scaling region defined in Section~\ref{sec:methodology} is obtained and the displayed parameter domain corresponds to the intersection of these validated regions. The standard kinematic decay-width constraint ($\Gamma/m_{X} = 15\%$) is omitted from the subplots as it lies entirely outside this reliable simulation domain across all benchmark masses.}
  \label{fig:excl_three_panels}
\end{figure*}

Figure \ref{fig:excl_three_panels} displays the overlaid 95\% CL exclusion contours in the $(\lambda_{1}, \lambda_{2})$ parameter space for the benchmark mediator masses $m_{X} = 1.5$, $1.75$, and $2.0~\text{TeV}$.

From the reinterpretation of the ATLAS $b$-tagged dijet resonance search, we obtain the exclusion bounds represented by the solid blue curves.
This dijet search exhibits high sensitivity primarily to larger $\lambda_{1}$ values, effectively probing the hadronic decay modes $X \to \bar{d}\bar{b}$ and $X \to \bar{s}\bar{b}$ of the scalar mediator.
For all three benchmark masses, the $b$-tagged dijet reinterpretation excludes $\lambda_1 > 0.45$ throughout the displayed validated parameter domains.
This common excluded region provides a conservative summary of the sensitivity across the three benchmarks, and the individual contours also exclude smaller values of $\lambda_1$ over portions of the parameter space.
A quantitative comparison was performed against the inclusive dijet channel for the $q^{*}$ model in the same ATLAS search~\cite{ATLAS:2019fgd}.
For $m_X = 2.0~\mathrm{TeV}$, the inclusive and $b$-tagged dijet reinterpretations yield comparable cross-section upper limits of approximately $0.062~\mathrm{pb}$ and $0.065~\mathrm{pb}$, respectively.
These are obtained from the reported $A \sigma$ limits using the corresponding \delphesShort signal acceptances.
The validated scaling region defined in Section~\ref{sec:methodology} is obtained for this channel, where the semi-analytical framework accurately fits the \mg simulation (relative difference $\leq 10\%$).

To assess whether the relative sensitivity of the two dijet strategies could change with increasing integrated luminosity, we additionally perform a simple luminosity-scaling estimate.
The cross-section upper limit is parameterized as $\sigma_{\mathrm{lim}} (L) = \sqrt{ A^{2} / L + S^{2} }$, where the two terms represent statistical and luminosity-independent systematic contributions, respectively.
For each analysis, $A$ and $S$ are determined from two published limits \cite{ATLAS:2018tfk, ATLAS:2019fgd, ATLAS:2017eqx} obtained for the same benchmark model at different integrated luminosities.
The resulting estimates indicate that the systematic contribution, $S^2$, remains substantially smaller than the statistical contribution, $A^2/L$, over the luminosity range considered for both channels.
Consequently, both limits continue to improve approximately with increasing luminosity, and the $b$-tagged dijet channel is not expected to outperform the inclusive dijet channel under this extrapolation.
Nevertheless, the increase in luminosity still improves the absolute sensitivity: for the $b$-tagged channel at $m_X=2.0~\text{TeV}$, the cross-section upper limit decreases from approximately $0.142~\text{pb}$ at $36.1~\text{fb}^{-1}$ \cite{ATLAS:2018tfk} to $0.065~\text{pb}$ at $139~\text{fb}^{-1}$, corresponding to an improvement of more than a factor of two.

Although the $b$-tagged dijet reinterpretation does not outperform the inclusive dijet reinterpretation, its relative sensitivity might be improved by exploiting the suppressed $b$-quark PDF.
For high mass resonances that require an incoming $b$ quark, this can favor configurations in which the $b$ quark originates from gluon splitting, potentially producing an additional $b$-tagged subleading jet as a characteristic signature.
Exploring this possibility, we further consider a modified selection requiring exactly one $b$-tagged jet in the leading dijet pair together with at least one additional $b$-tagged subleading jet.
For $m_{X}=2.0~\text{TeV}$, this requirement reduces the signal acceptance to approximately $36\%$ of that obtained with the nominal $\geq 1b$ leading-pair selection.
The corresponding QCD background suppression is estimated using $H_T$-binned multijet simulations, whose reconstructed $m_{jj}$ spectrum is found to be qualitatively consistent with the background shape reported by ATLAS.
With this validation, we consider the background ratio $B_{\mathrm{new}}/B_{\mathrm{nominal}}$ between the two selections.
Only a few scattered $m_{jj}$ bins yield $\sqrt{B_{\mathrm{new}}/B_{\mathrm{nominal}}}$ slightly below the signal-retention factor of approximately $0.35$, while the ratio remains above this threshold in most bins, including much of the region relevant to the $m_X=2.0~\text{TeV}$ benchmark.
Within the simple statistics-dominated estimate $S/\sqrt{B}$, the additional background suppression is therefore generally insufficient to compensate for the reduced signal acceptance under the nominal $b$-tagging working point.
Out of the remaining background, around 40\% stems from events with at least one jet misidentified as a $b$-jet.
This motivates emulating a tighter $b$-tagging working point, configured as a relative reduction of 10\% in true $b$ acceptance and a factor 10 reduction of the misidentification rate, not accounting for events with dual mistags as a conservative scenario.
Under these assumptions, the additional-$b$-tag selection can improve $S/\sqrt{B}$ relative to the nominal selection for $m_{jj} \gtrsim 1.6~\mathrm{TeV}$.
This scenario demonstrates that a dedicated search strategy based upon a second $b$ tag on a jet beyond the leading jet pair is, given proper optimization of or advances in $b$ tagging, feasible.

The reinterpretation of the ATLAS search for multijet final states with $p_{\text{T}}^{\text{miss}}$ (bottom squark) provides complementary constraints, represented by the solid red curves.
This channel is highly sensitive to the $X \to t \psi$ decay mode when the production coupling ($\lambda_1$) and the decay coupling ($\lambda_2$) are simultaneously large.
As shown in Figure \ref{fig:excl_three_panels}, the exclusion contour restricts the top-right corner of the parameter space, demonstrating that the maximum allowed $\lambda_2$ coupling decreases as $\lambda_1$ increases.
Across all tested masses, the bottom-squark reinterpretation excludes the common region with $\lambda_1 > 0.3$ and $\lambda_2 > 0.6$, restricted to the displayed parameter domains.
This channel also extends the exclusion to smaller values of $\lambda_1$ when $\lambda_2$ is sufficiently large.
For this topology, we also compare the SUSY multijet reinterpretation with a dedicated monotop search~\cite{ATLAS:2018cjd}.
This monotop analysis considers a scalar mediator $\phi$ and a DM particle $\chi$ with $m_\chi = 10~\mathrm{GeV}$.
At a common mediator mass of $m_\phi = m_X = 2.0~\mathrm{TeV}$, its 95\% CL cross-section upper limit is approximately one order of magnitude stronger than that obtained from our SUSY reinterpretation.
Nevertheless, the two analyses target different event configurations and may select largely complementary samples.
If their signal regions are shown to be orthogonal, or if any overlap is properly modeled, their likelihoods could be combined with correlated systematic uncertainties treated consistently, potentially strengthening the overall constraint.
Similar to the $b$-tagged dijet analysis, an independent validated scaling region is derived for this channel to identify where its respective semi-analytical scaling meets the 10\% accuracy criterion.

In the final exclusion plots, we display only the intersection of the reliable simulation regions obtained from the $b$-tagged dijet and bottom squark studies, ensuring that both reinterpretations are shown only where their respective scaling relations reproduce the \mg results within the specified accuracy.
It should be noted that a theoretical decay-width constraint ($\Gamma/m_X = 15\%$), beyond which the narrow-width approximation breaks down, applies universally to both the $b$-tagged dijet and SUSY channels for a given mediator mass.
For a benchmark mass of $m_X = 1.5~\text{TeV}$, this boundary passes through the coupling values $\lambda_1 \approx 1.3$ and $\lambda_2 \approx 0.9$.
At this point, the large-width and off-shell effects cause the \mg simulated cross sections to deviate from the semi-analytical narrow-width scaling by 18\% in the dijet channel and 9.4\% in the bottom squark channel.
Consequently, this constraint is not displayed in the final exclusion figures, as it lies entirely outside the common reliable simulation region defined by the intersection of the $b$-tagged dijet and bottom squark validity domains.

Finally, we evaluated the reinterpretation of the ATLAS search targeting top squarks, characterized by $\tau$ leptons, $b$-jets, and $p_{\text{T}}^{\text{miss}}$ originating from the leptonic decay of the top quark.
However, this channel did not yield competitive exclusion bounds.
The ATLAS analysis utilizes a multivariate, recurrent neural-network-based $\tau$-tagging algorithm \cite{ATLAS:2019uhp} calibrated to the Medium working point guaranteeing approximately 75\% identification efficiency for hadronic one-prong $\tau$ decays (60\% for three-prong decays).
In our fast-simulation setup, precisely reproducing the performance of such highly optimized, detector-specific machine learning algorithms is inherently challenging.
By evaluating truth-matched generator-level taus against the reconstructed objects in our simulated signal sample, the effective $\tau$-tagging efficiency achieved was 22.75\% ($1462/6427$).
Because the cumulative signal acceptance was consequently much lower than the experimental expectation, the predicted yields were insufficient to derive meaningful constraints in the $\lambda_1$-$\lambda_2$ parameter plane.
As a result, exclusion contours for this channel are not presented.

\section{Summary} \label{sec:summary}

In this work, we presented a reinterpretation of ATLAS Run 2 searches to probe a $\Delta B = 2$ BNV model (the mono-$X$ model).
This framework introduces scalar mediators that couple to SM quarks and a fermionic DM candidate, $\psi$.
Depending on the relative strengths of the couplings $\lambda_1$ and $\lambda_2$, the lighter mediator $X$ predominantly decays either into a pair of down-type quarks ($\bar{d}\bar{b}$ or $\bar{s}\bar{b}$) or into a top quark and missing transverse momentum ($X \to t\psi$).
By mapping these theoretical signatures onto existing ATLAS analyses targeting $b$-tagged dijet resonances and multijet plus $p_{\text{T}}^{\text{miss}}$ final states, we derived new 95\% CL exclusion limits on the model parameter space.

Our findings demonstrate that existing experimental searches are highly complementary.
The $b$-tagged dijet resonance search provides stringent constraints on scenarios dominated by the $\lambda_1$ coupling, effectively probing the purely hadronic decay modes of the mediator.
Conversely, the reinterpretation of the bottom squark search yields robust constraints when $\lambda_1$ (which drives the production cross section) and $\lambda_2$ (which governs the decay) are both large.
For benchmark mediator masses of 1.5, 1.75, and 2.0 TeV, we mapped the excluded regions in the $\lambda_1$-$\lambda_2$ parameter plane within the domain where the semi-analytical cross sections agree with the simulated cross sections to within 10\%.
A preliminary study further indicates that an additional subleading $b$-tagged jet could improve sensitivity near $m_{jj}\sim2~\mathrm{TeV}$ under the assumed tighter tagging working point, though the nominal working point generally does not improve sensitivity.

Additionally, we explored the reinterpretation of the stop search, which targets final states with $\tau$ leptons, $b$-jets, and $p_{\text{T}}^{\text{miss}}$.
However, this channel proved inconclusive in our study.
The effective $\tau$ identification efficiency obtained in our simulation was substantially lower than the performance of the specialized $\tau$-tagging algorithms employed by the ATLAS collaboration.
This discrepancy resulted in a highly conservative estimate of the signal acceptance, which suppressed the predicted yields and prevented the derivation of reliable limits.
This observation reflects the general difficulty of precisely reproducing advanced, detector-specific machine learning calibrations in phenomenological reinterpretations.

By reinterpreting existing ATLAS measurements in the context of the mono-$X$ framework, this work establishes new constraints on the mono-$X$ model parameter space.
These constraints delineate the excluded regions for the evaluated mediator masses and couplings, while explicitly identifying the remaining parameter space that can be further probed with the increased statistics of the High-Luminosity LHC.

\paragraph*{Acknowledgement}
The authors gratefully acknowledge support from Vanderbilt University and the U.S. National Science Foundation.
This work is supported in part by NSF Award PHY-2411502.

%% If you have bibdatabase file and want bibtex to generate the
%% bibitems, please use
%%
\bibliographystyle{elsarticle-num}
\biboptions{numbers,sort&compress}
\bibliography{refs}

@article{Canetti:2012zc,
    author = "Canetti, Laurent and Drewes, Marco and Shaposhnikov, Mikhail",
    title = "{Matter and Antimatter in the Universe}",
    eprint = "1204.4186",
    archivePrefix = "arXiv",
    primaryClass = "hep-ph",
    reportNumber = "TTK-12-04",
    doi = "10.1088/1367-2630/14/9/095012",
    journal = "New J. Phys.",
    volume = "14",
    pages = "095012",
    year = "2012"
}

@article{Kalsi:2024hfe,
    author = "Kalsi, Amandeep Kaur and Kamon, Teruki and Kim, Seulgi and Lee, Jason S. H. and Rathjens, Denis and Roh, Youn Jung and Thompson, Adrian and Watson, Ian James",
    title = "{Probing bottom-associated production of a TeV scale scalar decaying to a top quark and dark matter at the LHC}",
    eprint = "2404.14844",
    archivePrefix = "arXiv",
    primaryClass = "hep-ph",
    doi = "10.1007/JHEP09(2024)203",
    journal = "JHEP",
    volume = "09",
    pages = "203",
    year = "2024"
}

@article{ATLAS:2021jyv,
    author = "Aad, Georges and others",
    collaboration = "ATLAS",
    title = "{Search for new phenomena in $pp$ collisions in final states with tau leptons, b-jets, and missing transverse momentum with the ATLAS detector}",
    eprint = "2108.07665",
    archivePrefix = "arXiv",
    primaryClass = "hep-ex",
    reportNumber = "CERN-EP-2021-118",
    doi = "10.1103/PhysRevD.104.112005",
    journal = "Phys. Rev. D",
    volume = "104",
    number = "11",
    pages = "112005",
    year = "2021"
}

@article{ATLAS:2021yij,
    author = "Aad, Georges and others",
    collaboration = "ATLAS",
    title = "{Search for new phenomena in final states with $b$-jets and missing transverse momentum in $\sqrt{s}=13$ TeV $pp$ collisions with the ATLAS detector}",
    eprint = "2101.12527",
    archivePrefix = "arXiv",
    primaryClass = "hep-ex",
    reportNumber = "CERN-EP-2021-001",
    doi = "10.1007/JHEP05(2021)093",
    journal = "JHEP",
    volume = "05",
    pages = "093",
    year = "2021"
}

@article{ATLAS:2018tfk,
    author = "Aaboud, Morad and others",
    collaboration = "ATLAS",
    title = "{Search for resonances in the mass distribution of jet pairs with one or two jets identified as $b$-jets in proton-proton collisions at $\sqrt{s}=13$ TeV with the ATLAS detector}",
    eprint = "1805.09299",
    archivePrefix = "arXiv",
    primaryClass = "hep-ex",
    reportNumber = "CERN-EP-2018-075",
    doi = "10.1103/PhysRevD.98.032016",
    journal = "Phys. Rev. D",
    volume = "98",
    pages = "032016",
    year = "2018"
}

@article{Alwall:2014hca,
    author = "Alwall, J. and Frederix, R. and Frixione, S. and Hirschi, V. and Maltoni, F. and Mattelaer, O. and Shao, H. -S. and Stelzer, T. and Torrielli, P. and Zaro, M.",
    title = "{The automated computation of tree-level and next-to-leading order differential cross sections, and their matching to parton shower simulations}",
    eprint = "1405.0301",
    archivePrefix = "arXiv",
    primaryClass = "hep-ph",
    reportNumber = "CERN-PH-TH-2014-064, CP3-14-18, LPN14-066, MCNET-14-09, ZU-TH-14-14",
    doi = "10.1007/JHEP07(2014)079",
    journal = "JHEP",
    volume = "07",
    pages = "079",
    year = "2014"
}

@article{Sjostrand:2014zea,
    author = {Sj{\"o}strand, Torbj{\"o}rn and Ask, Stefan and Christiansen, Jesper R. and Corke, Richard and Desai, Nishita and Ilten, Philip and Mrenna, Stephen and Prestel, Stefan and Rasmussen, Christine O. and Skands, Peter Z.},
    title = "{An introduction to PYTHIA 8.2}",
    eprint = "1410.3012",
    archivePrefix = "arXiv",
    primaryClass = "hep-ph",
    reportNumber = "LU-TP-14-36, MCNET-14-22, CERN-PH-TH-2014-190, FERMILAB-PUB-14-316-CD, DESY-14-178, SLAC-PUB-16122",
    doi = "10.1016/j.cpc.2015.01.024",
    journal = "Comput. Phys. Commun.",
    volume = "191",
    pages = "159--177",
    year = "2015"
}

@article{deFavereau:2013fsa,
    author = "de Favereau, J. and Delaere, C. and Demin, P. and Giammanco, A. and Lema{\^\i}tre, V. and Mertens, A. and Selvaggi, M.",
    collaboration = "DELPHES 3",
    title = "{DELPHES 3, A modular framework for fast simulation of a generic collider experiment}",
    eprint = "1307.6346",
    archivePrefix = "arXiv",
    primaryClass = "hep-ex",
    doi = "10.1007/JHEP02(2014)057",
    journal = "JHEP",
    volume = "02",
    pages = "057",
    year = "2014"
}

@article{2013290,
title = {Parton distributions with QED corrections},
journal = {Nuclear Physics B},
volume = {877},
number = {2},
pages = {290-320},
year = {2013},
issn = {0550-3213},
doi = {10.1016/j.nuclphysb.2013.10.010},
url = {https://www.sciencedirect.com/science/article/pii/S0550321313005233},
author = {Richard D. Ball and Valerio Bertone and Stefano Carrazza and Luigi {Del Debbio} and Stefano Forte and Alberto Guffanti and Nathan P. Hartland and Juan Rojo}
}

@article{Skands:2014pea,
    author = "Skands, Peter and Carrazza, Stefano and Rojo, Juan",
    title = "{Tuning PYTHIA 8.1: the Monash 2013 Tune}",
    eprint = "1404.5630",
    archivePrefix = "arXiv",
    primaryClass = "hep-ph",
    reportNumber = "CERN-PH-TH-2014-069, MCNET-14-08, OUTP-14-05P",
    doi = "10.1140/epjc/s10052-014-3024-y",
    journal = "Eur. Phys. J. C",
    volume = "74",
    number = "8",
    pages = "3024",
    year = "2014"
}

@article{ATLAS:2019uhp,
    author = {{ATLAS Collaboration}},
    collaboration = "ATLAS",
    title = "{Identification of hadronic tau lepton decays using neural networks in the ATLAS experiment}",
    journal = "Report No. ATL-PHYS-PUB-2019-033",
    reportNumber = "ATL-PHYS-PUB-2019-033",
    year = "2019"
}

@article{Planck:2018vyg,
    author = "Aghanim, N. and others",
    collaboration = "Planck",
    title = "{Planck 2018 results. VI. Cosmological parameters}",
    eprint = "1807.06209",
    archivePrefix = "arXiv",
    primaryClass = "astro-ph.CO",
    doi = "10.1051/0004-6361/201833910",
    journal = "Astron. Astrophys.",
    volume = "641",
    pages = "A6",
    year = "2020",
    note = "[Erratum: Astron.Astrophys. 652, C4 (2021)]"
}

@article{
    CMS:2024onh,
    author = "Hayrapetyan, Aram and others",
    collaboration = "CMS",
    title = "The {CMS} statistical analysis and combination tool: {\textsc{Combine}}",
    eprint = "2404.06614",
    archivePrefix = "arXiv",
    primaryClass = "physics.data-an",
    reportNumber = "CMS-CAT-23-001, CERN-EP-2024-078",
    year = "2024",
    journal = "Comput. Softw. Big Sci.",
    doi = "10.1007/s41781-024-00121-4",
    volume = "8",
    pages = "19"
}

@article{Mangano:2006rw,
    author = "Mangano, Michelangelo L. and Moretti, Mauro and Piccinini, Fulvio and Treccani, Michele",
    title = "{Matching matrix elements and shower evolution for top-quark production in hadronic collisions}",
    eprint = "hep-ph/0611129",
    archivePrefix = "arXiv",
    reportNumber = "CERN-PH-TH-2006-232, FNT-T-2006-09",
    doi = "10.1088/1126-6708/2007/01/013",
    journal = "JHEP",
    volume = "01",
    pages = "013",
    year = "2007"
}

@article{Alwall:2007fs,
    author = "Alwall, Johan and others",
    title = "{Comparative study of various algorithms for the merging of parton showers and matrix elements in hadronic collisions}",
    eprint = "0706.2569",
    archivePrefix = "arXiv",
    primaryClass = "hep-ph",
    reportNumber = "SLAC-PUB-12604, CERN-PH-TH-2007-066, LU-TP-07-13, KA-TP-06-2007, DCPT-07-62, IPPP-07-31",
    doi = "10.1140/epjc/s10052-007-0490-5",
    journal = "Eur. Phys. J. C",
    volume = "53",
    pages = "473--500",
    year = "2008"
}

@article{ATLAS:2019fgd,
    author = "Aad, Georges and others",
    collaboration = "ATLAS",
    title = "{Search for new resonances in mass distributions of jet pairs using 139 fb$^{-1}$ of $pp$ collisions at $\sqrt{s}=13$ TeV with the ATLAS detector}",
    eprint = "1910.08447",
    archivePrefix = "arXiv",
    primaryClass = "hep-ex",
    reportNumber = "CERN-EP-2019-162",
    doi = "10.1007/JHEP03(2020)145",
    journal = "JHEP",
    volume = "03",
    pages = "145",
    year = "2020"
}

@article{ATLAS:2018cjd,
    author = "Aaboud, Morad and others",
    collaboration = "ATLAS",
    title = "{Search for large missing transverse momentum in association with one top-quark in proton-proton collisions at $ \sqrt{s} $ = 13 TeV with the ATLAS detector}",
    eprint = "1812.09743",
    archivePrefix = "arXiv",
    primaryClass = "hep-ex",
    reportNumber = "CERN-EP-2018-301",
    doi = "10.1007/JHEP05(2019)041",
    journal = "JHEP",
    volume = "05",
    pages = "041",
    year = "2019"
}

@article{ATLAS:2017eqx,
    author = "Aaboud, Morad and others",
    collaboration = "ATLAS",
    title = "{Search for new phenomena in dijet events using 37 fb$^{-1}$ of $pp$ collision data collected at $\sqrt{s}=$13 TeV with the ATLAS detector}",
    eprint = "1703.09127",
    archivePrefix = "arXiv",
    primaryClass = "hep-ex",
    reportNumber = "CERN-EP-2017-042",
    doi = "10.1103/PhysRevD.96.052004",
    journal = "Phys. Rev. D",
    volume = "96",
    number = "5",
    pages = "052004",
    year = "2017"
}

@article{CMS:2022eud,
    author = "Tumasyan, Armen and others",
    collaboration = "CMS",
    title = "{Search for narrow resonances in the b-tagged dijet mass spectrum in proton-proton collisions at $\sqrt{s}=13~\mathrm{TeV}$}",
    eprint = "2205.01835",
    archivePrefix = "arXiv",
    primaryClass = "hep-ex",
    reportNumber = "CMS-EXO-20-008, CERN-EP-2022-037",
    doi = "10.1103/PhysRevD.108.012009",
    journal = "Phys. Rev. D",
    volume = "108",
    number = "1",
    pages = "012009",
    year = "2023"
}

@article{CMS:2019gwf,
    author = "Sirunyan, Albert M and others",
    collaboration = "CMS",
    title = "{Search for high mass dijet resonances with a new background prediction method in proton-proton collisions at $\sqrt{s} =$ 13 TeV}",
    eprint = "1911.03947",
    archivePrefix = "arXiv",
    primaryClass = "hep-ex",
    reportNumber = "CMS-EXO-19-012, CERN-EP-2019-222",
    doi = "10.1007/JHEP05(2020)033",
    journal = "JHEP",
    volume = "05",
    pages = "033",
    year = "2020"
}

@article{ATLAS:2022qxm,
    author = "Aad, Georges and others",
    collaboration = "ATLAS",
    title = "{ATLAS flavour-tagging algorithms for the LHC Run 2 pp collision dataset}",
    eprint = "2211.16345",
    archivePrefix = "arXiv",
    primaryClass = "physics.data-an",
    reportNumber = "CERN-EP-2022-226",
    doi = "10.1140/epjc/s10052-023-11699-1",
    journal = "Eur. Phys. J. C",
    volume = "83",
    number = "7",
    pages = "681",
    year = "2023"
}

%% else use the following coding to input the bibitems directly in the
%% TeX file.

%%\begin{thebibliography}{00}

%% \bibitem[Author(year)]{label}
%% For example:

%% \bibitem[Aladro et al.(2015)]{Aladro15} Aladro, R., Martín, S., Riquelme, D., et al. 2015, \aas, 579, A101f

%%\end{thebibliography}

\end{document}